\documentclass[reprint,aps,pra]{revtex4-1}

\usepackage{graphicx}
\usepackage{epstopdf}
\usepackage{braket}

\begin{document}
\draft

\title{
Convergence of transition amplitudes obtained with the \\
Schwinger variational principle
}

\vspace{1cm}

\author{
 V. D. Rodr\'{\i}guez}
\affiliation{Departamento de F\'{\i}sica and IFIBA CONICET, FCEyN, UBA, Ciudad Universitaria Pabellón 1, 1428 Buenos Aires, Argentina}
\email[E-mail me at: ]{vladimir@df.uba.ar}


\begin{abstract}
An exactly solvable  time-dependent  quantum mechanical problem is employed
to study the convergence  properties of transition amplitudes calculated by
using the Schwinger variational principle.    A  detailed comparison
between the
amplitudes  approximated by  the  perturbative series  and  by
their  associated
Schwinger variational principles is performed.  The much better performance
obtained  by the variational principle is documented  through  different
case studies.
  For a given  order  of the Schwinger principle,
  it is observed that
the  transition  amplitudes  do  not  converge to the exact one  for  large
perturbations. The latter is true
 even though
large
combinations  of  unperturbed  states with  constant
coefficients
are taken  as trial wave  functions.
 As a matter of fact, it  is shown that
 the improvement of the method comes from
using  better  trial  wave functions and
increasing the order of the Schwinger principle employed.
\end{abstract}

\maketitle

\narrowtext
\section{INTRODUCTION}

Time  dependent  quantum  systems  arise  from  different branches of  atomic
physics,   such as atomic  collisions  or  strong-laser  atom
interaction studies.  Some important efforts to
bring to light the comprehension
of  these systems
beyond  the  well-known  adiabatic  or  perturbative  limits  consist  on
attempting to solve numerically
the time-dependent Schr\"odinger
equation        governing                the        system        evolution
\cite{Fritsch91,McCurdy91,Rodriguez92,Martin95}.

When dealing with
 atomic collision problems, alternative methods were proposed
to cope with
 atomic  excitation  (ionization)  by  fast highly charged ions.
 For  instance,  the  so  called    distorted    wave    methods:
 symmetric
eikonal (SE)
\cite{Deco86},
the continuum distorted wave - eikonal initial state
CDW-EIS \cite{Crothers83}, and the  eikonal-impulse approximation
(EIA) \cite{Rodriguez90}
 were found to be
 successful to explain experimental results.

On the other  hand,   the
Schwinger variational principle
was also able to reproduce experimental
data    of  Fe$^+24$  excitation  by  neutral  gases\cite{Gayet85,Gayet89}.
 However, difficulties  associated  with the  calculation
 of the required second order
Born approximation
 make it hard to achieve
a  complete  assessment  of  this theory.
This, in spite of recent work to include the whole discrete spectrum into the
closure  calculation \cite{Gayet89}.
 In a different approach, Martin and Salin have recently
 obtained the second order Born approximation by relating it
 to the close-coupling calculations.  Nevertheless, the accuracy of the
 two  procedures  mentioned require some further test, as also admitted
 by the authors.
 Moreover,  the    calculations    of
variational functionals involving Born terms of  higher  order than two
remain
at present  a non-tractable evaluation  due to the large amount of
computational task demanded.

In addition,  from  a theoretical point of view it is always of interest to
test variational principles,  such  as the Schwinger variational principle,
that do not represent minimum principles.  The amplitudes obtained with the
Schwinger variational principle cannot be employed to bound the exact ones,
but just to state that if  one  is  close  enough  to  the  scattering wave
function, then the resulting transition amplitude differs  from  the  exact
one in a second order quantity \cite{Dettmann72}.

In  this  paper  we  plan  to study  the  convergence of  the  variational
functional as applied to an exactly solvable model  representing  a distant
collision  problem.
 We require the model to have analytical transition amplitudes in order to
 generate the whole perturbative series, and also to be able
to use it as a benchmark
 to test the Schwinger variational principles.
  In  section II, different  orders  of  the
Schwinger variational  transition
amplitudes for a time dependent problem are obtained.  Section III
is devoted to present the model in the context of ion-atom excitation.  The
results obtained by using the Schwinger functionals
   are  presented  in  section  IV,  and  a  comparison
with  the
corresponding perturbative  results  as  well  as with
 the exact results is
performed.  Finally, conclusions
are drawn in section V.
Atomic units will be employed along this work.

\section{SCHWINGER VARIATIONAL AMPLITUDES FOR
               TIME DEPENDENT PROBLEMS}

Let us consider a system where the
 time dependent Hamiltonian $H$ can be separated
into a stationary part $H_0$ characterizing  the  unperturbed  system and a
time-dependent interaction potential $V({\bf r},t)$.  We  consider  a
one electron atomic hamiltonian
\begin{equation}
H_0 = -{\nabla^2_{\bf r} \over 2} + V_T(r)
\end{equation}
where $V_T(r)$ is the atomic interaction.

In  what  follows,  we use  the interaction  picture,  in  which  the
perturbing potential is given by
\begin{equation}
V(t)= \exp(i H_0 t) V({\bf r},t)  \exp(-i H_0 t)
\end{equation}
 The  dependence on the electron coordinate ${\bf r}$ was  dropped  to
distinguish  the change
from the Scr\"odinger picture to the
Heisemberg picture.  Provided that $ V(t\to\pm \infty)\to 0  $,
the amplitude for the
transition  from an initial unperturbed state $\varphi_i$
to an excited state $\varphi_f$ is
\begin{equation}
a_{fi}=
\lim_{t\to +\infty}
 \braket{\varphi_f|\psi_i^+(t)}=
\lim_{t\to\ -\infty}
\braket{\psi_f^-(t)|\varphi_i}
\end{equation}
where  $\varphi_{i,f}$   are  stationary  solutions  of  the
unperturbed Schr\"odinger
equation given by
\begin{equation}
  H_0 \  \ket{\varphi_{i,f} }= \varepsilon_{i,f} \   \ket{\varphi_{i,f}}
\end{equation}
and the scattering  wave functions $\psi_i^+(t) $ and $\psi_f^-(t)$ satisfy
the Lippmann-Schwinger equations
\begin{eqnarray}
\ket{\psi_i^+(t)}&=& \ket{\varphi_i} -i\int_{-\infty}^{t}d t^{\prime}
                             V(t^{\prime})\ket{\psi_i^+(t^\prime)} \\
\ket{\psi_f^-(t)}&=& \ket{\varphi_f} -i\int_{t}^{+\infty}d t^{\prime}
                             V(t^{\prime})\ket{\psi_f^-(t^\prime)}
\end{eqnarray}

Note  that    $\ket{\psi_f^-(t)}$  is  the  advanced  scattering  state, {\it i.e.}
$ \ket{\psi_f^-(t)} \to \ket{\varphi_f}$ as
$t \to \infty$.

By introducing the Lippman Schwinger equations in  Eq.  3
we find the alternative forms for the transition amplitudes
\begin{eqnarray}
a_{fi}&=& \delta_{i,f}  -i\int_{-\infty}^{\infty} d t
                        \braket{\varphi_f| V(t)|\psi_i^+(t)} \\
      &=& \delta_{i,f}  -i\int_{-\infty}^{\infty} d t
                       \braket{\psi_f^-(t)| V(t)|\varphi_i}
\end{eqnarray}

As it is well known,
 the Lippmann-Schwinger Eqs. (5)  and  (6) may be iterated to produce the
$N$th order of the Born perturbative series
\begin{eqnarray}
a_{fi}^{(N)}&&= \delta_{i,f} - i\int_{-\infty}^{\infty} d t_1
               \braket{\varphi_f| V(t_1)|\varphi_i}  +  \nonumber \\
&& + (-i)^2 \int_{-\infty}^{\infty} d t_1 \int_{-\infty}^{t_1} d t_2
         \braket{\varphi_f| V(t_1)\ V(t_2)|\varphi_i} + ...  \nonumber\\
&& + (-i)^N \int_{-\infty}^{\infty} d t_1 \int_{-\infty}^{t_1} d t_2 ...
         \int_{-\infty}^{t_{N-1}} d t_N \nonumber\\
 &&        \braket{\varphi_f| V(t_1)\ V(t_2)..V(t_{N})|\varphi_i}
\end{eqnarray}

The  order  of  the Born approximation
is given by the number of
times the interaction potential is present
in the last term of the truncated series.

 We associate a Schwinger variational
functional  with  each one
of the  perturbative  expansion  orders of  the  transition
amplitude.
 This task has been done in the context of
 time independent collisions in reference \cite{Jorge82}.
 For  this  purpose,  we  generalize  the  procedure employed in
reference  \cite{Gayet89}.   We  rewrite  the  transition  amplitude  in  the
following equivalent ways

\begin{eqnarray}
a_{fi}=&& a_{fi}^{(N-2)} + (-i)^{N-1} \int_{-\infty}^{\infty} d t_1 \int_{-\infty}^{t_1} d t_2 ... \int_{-\infty}^{t_{N-2}} d t_{N-1}\nonumber \\
     &&  \braket{\varphi_f| V(t_1)\ V(t_2) ... V(t_{N-1}) |\psi_i^+(t_{N-1})}
\label{a}
\\
   =&& a_{fi}^{(N-2)} + (-i)^{N-1}
 \int_{-\infty}^{\infty} d t_1 \int_{t_1}^{\infty} d t_2 ...
   \int_{t_{N-2}}^{\infty} d t_{N-1}   \nonumber \\
    &&  \braket{\psi_f^-(t_{N-1})| V(t_{N-1})\ V(t_{N-2})...V(t_1)|\varphi_i}
\label{b}
\end{eqnarray}
where $a_{fi}^{(N-2)}$ denotes the $(N-2)$th order in the series expansion
in  Eq. 9.

Other equivalent expressions for the transition amplitude can  be generated
by replacing $\varphi_i$ and $\varphi_f$ in Eqs.  \ref{a}  and  \ref{b}  by
the forms obtained from Eqs.  7 and 8, respectively:

\begin{eqnarray}
a_{fi}=&& a_{fi}^{(N-2)} + (-i)^{N-1}
 \int_{-\infty}^{\infty} d t_1 \int_{t_1}^{\infty} d t_2 ...
   \int_{t_{N-2}}^{\infty} d t_{N-1}  \nonumber \\
 &&    \braket{\psi_f^-(t_{N-1})| V(t_{N-1})\ V(t_{N-1})...V(t_1)|\psi_i^+(t_1)} +
    \nonumber \\
&& + (-i)^N \int_{-\infty}^{\infty} d t_1 \int_{t_1}^{\infty} d t_2 ...
   \int_{t_{N-1}}^{\infty} d t_{N} \nonumber \\
&&    \braket{\psi_f^-(t_{N})| V(t_{N})\ V(t_{N-1})...V(t_1)|\psi_i^+(t_1)}
\\
  =&& a_{fi}^{(N-2)} + (-i)^{N-1}
 \int_{-\infty}^{\infty} d t_1 \int_{-\infty}^{t_1} d t_2 ...
   \int_{-\infty}^{t_{N-2}} d t_{N-1}  \nonumber \\
   && \braket{\psi_f^-(t_1)| V(t_1)\ V(t_2...V(t_{N-1})|\psi_i^+(t_{N-1})} +
    \nonumber \\
&& + (-i)^N \int_{-\infty}^{\infty} d t_1 \int_{-\infty}^{t_1} d t_2 ...
   \int_{-\infty}^{t_{N-1}} d t_{N} \nonumber \\
  &&  \braket{\psi_f^-(t_1)| V(t_1)\ V(t_2)...V(t_N)|\psi_i^+(t_N)}
\end{eqnarray}

Finally, we can generate  other alternative expressions for the amplitude
 by performing Eq. 10 + Eq. 11 - Eq. 12, as follows:
\begin{eqnarray}
a_{fi}=&&a_{fi}^{(N-2)} + (-i)^{N-1}
 \int_{-\infty}^{\infty} d t_1 \int_{-\infty}^{t_1} d t_2 ...
   \int_{-\infty}^{t_{N-2}} d t_{N-1}\nonumber \\
  &&  \braket{\varphi_f| V(t_1)\ V(t_2) ... V(t_{N-1}) |\psi_i^+(t_{N-1})} +
     \nonumber \\
&& + (-i)^{N-1}
 \int_{-\infty}^{\infty} d t_1 \int_{t_1}^{\infty} d t_2 ...
   \int_{t_{N-2}}^{\infty} d t_{N-1} \nonumber \\
&&    \braket{\psi_f^-(t_{N-1})| V(t_{N-1})\ V(t_{N-2})...V(t_1)|\varphi_i} +
    \nonumber  \\
&& (-i)^{N-1}
 \int_{-\infty}^{\infty} d t_1 \int_{t_1}^{\infty} d t_2 ...
   \int_{t_{N-2}}^{\infty} d t_{N-1}\nonumber \\
   &&  \braket{\psi_f^-(t_{N-1})| V(t_{N-1})\ V(t_{N-1})...V(t_1)|\psi_i^+(t_1)} +
     \nonumber \\
&& (-i)^N \int_{-\infty}^{\infty} d t_1 \int_{t_1}^{\infty} d t_2 ...
   \int_{t_{N-1}}^{\infty} d t_{N} \nonumber \\
  &&   \braket{\psi_f^-(t_{N})| V(t_{N})\ V(t_{N-1})...V(t_1)|\psi_i^+(t_1)}
\end{eqnarray}

This  expression  is  the  Schwinger  functional we were looking for.    By
construction,  it  gives  the exact transition amplitude by using the exact
wave functions.  Furtherly, it is not difficult to show that this  functional
is stationary for small variations around the exact wave functions.  It
may be also  demonstrated  that  when  unperturbed  wave functions are employed
instead of the scattering wave functions, the perturbative transition
amplitude to $n$th order is obtained.
   As  an alternative derivation, we might perform
Eq.  10 + Eq.  11 - Eq.  13  to obtain the same variational functional.

In order to employ the variational principle, we expand the scattering  wave
functions $\ket{\psi_i^+}$ and $\ket{\psi_f^-}$ on a truncated basis sets \{$\ket{m}$\} and
\{$\ket{n}$\}, respectively.    The  expansion coefficients can be determined by
using the variational condition $\delta a_{fi}=0$.  By
solving a linear system,
 we arrive at the following variationally determined transition amplitude

\begin{equation}
a_{fi}^{(N)}= a_{fi}^{(N-2)} + {\bf V}^{(N-1)}.({\bf D}^{(N)})^{-1}.
         {\bf V}^{(N-1)}
\end{equation}
where the matrix elements of ${\bf  V}^{(N-1)}$  and  ${\bf  D}^{(N)}$  are
given by
\begin{equation}
{\bf  V}_{mn}^{(N-1)} = a_{mn}^{(N-1)} -a_{mn}^{(N-2)}\end{equation}
and
\begin{equation}
{\bf  D}_{mn}^{(N-1)} = {\bf  V}_{mn}^{(N-1)} - {\bf  V}_{mn}^{(N)}
\end{equation}
It should be noted that  ${\bf  V}_{mn}^{(N-1)}$ is nothing but the
$(N-1)$th order term in the Born expansion of the transition amplitude.

Equation 15 and 16  may be considered as the generalization of equations
14 and 15  in reference  \cite{Gayet89},  where only
the second order $N=2$ and  the inelastic case were presented.

We remark that, even though the basis sets \{$\ket{m}$\} and \{$\ket{n}$\} are
complete, we should not expect the resulting variational amplitude to be
equal to the exact one.   This  is so because a limited class of trial wave
functions  was employed, namely that represented by  a  linear  combination
with time-independent coefficients.

\section{THE DISTANT COLLISION MODEL}

In this  section,  we  summarize the model that is employed to test the
convergence of the  Schwinger  variational  functional.    Let
us consider the
collisions of a heavy  fast  particle  of  charge $Z_P$ interacting with an
electron bounded to an `atom'  that  is  represented by an isotropic harmonic
oscillator, whose target potential is given by
\begin{equation}
V_T(r) = {w^2 \over  2} (x^2+y^2+z^2)
\end{equation}
The projectile follows a classical trajectory
\begin{equation}
{\bf R}(t) =
{\bf b} + {\bf v}t
\end{equation}
determined  by the impact parameter ${\bf
b}$,  and  the impact velocity ${\bf v}$.   We  consider   distant
collisions, {\it i.e.},  large  impact parameters as compared
with the mean  atomic radio. Under this condition,
the Coulomb interaction between the projectile
and the electron can be safely represented by its dipolar approximation
\begin{equation}
V({\bf r},t)=-Z_P {{\bf R}(t) \cdot {\bf  r}  \over  R(t)^3}
\end{equation}
where the monopolar term has  been  omitted,  since  it does not affect the
transition probabilities.

The  Hamiltonian  of  the  system  above
corresponds to the one of  a  forced  isotropic
oscillator. This model
was    previously    employed    by    Hill    and
Merzbacher\cite{Merzbacher74} as a  starting point for polarization studies
in Coulomb collisions of  charged  particles  with  atoms.
 The evolution
operator of this time dependent  quantum system  may be exactly obtained by
separation of variables in the cartesian  coordinates $x,y,z$.

We take the  impact  parameter  ${\bf b}$ along the $y$-axes, and the impact
velocity  ${\bf  v}$ in  the  $z$  direction,  {\it  i.e.},  there  are  no
perturbations in the $x$-direction.  Thus,
 the transition amplitude to go
from  the  unperturbed  initial  state  $\ket{0,n_y,n_z}$   to  a  final  state
$\ket{0,m_y,m_z}$,
in terms of one-dimensional transition amplitudes $a(\alpha,\beta,n,m)$,
is given by
\begin{equation}
a_{\{m\},\{n\}}= a(\alpha_y,\beta_y,n_y,m_y) a(\alpha_z,\beta_z,n_z,m_z)
\end{equation}
 These  amplitudes  were
 obtained  in  different  ways;  we  refer  the  reader  to
\cite{Pechukas66}  for  the analytical
 calculations. Here, we just  quote  the  required
results. Provided that the forcing term $V(q,t)$ is expressed as
\begin{equation}
V(q,t)= (2w) f(t)
\end{equation}
the one-dimensional transition amplitudes \cite{Pechukas66} are given by
\begin{eqnarray}
a(\alpha,\beta,n,m)  =& \exp(i \beta - |\alpha|^2/2)
(m!)^{1\over 2} (n!)^{1\over 2} (i \alpha^*)^{n-m} \times \nonumber \\
&\sum_{j=0}^{m}  {  (-)^j  |\alpha|^{2j}  \over  j!    (n-m+j)!   (m-j)!  } \ \ \ \ n \ge m  \\
=& \exp(i \beta - |\alpha|^2/2)
(m!)^{1\over 2} (n!)^{1\over 2} (i \alpha)^{m-n} \times \nonumber \\
&\sum_{k=0}^{n} { (-)^k |\alpha|^{2k} \over  k! (m-n+k)! (n-k)! }
\ \ \ \ n \le m
\end{eqnarray}
where the magnitudes $\alpha$ and $\beta$ are defined as
\begin{equation}
 \alpha = -\int_{-\infty}^{\infty} d t' \exp(i w t') f(t')
\end{equation}
\begin{equation}
 \beta = -\int_{-\infty}^{\infty} d t_2
 \int_{-\infty}^{t_2} d t_1 f(t_1) f(t_2) \sin[w(t_1-t_2)]
\end{equation}

In the present  case  $\alpha_{y,z}$  have  closed  forms  in  terms of the
modified Bessel functions $K_1$ and $K_0$:
\begin{eqnarray}
\alpha_y =& {2 Z_P \over b v} |a|K_1(|a|) {1\over \sqrt{2w}}  \\
\alpha_z =& {2 i Z_P \over b v} a K_0(|a|) {1\over \sqrt{2w}}  \\
\end{eqnarray}
where the parameter $a=w b/v $.
On the other hand,  the  calculation  of  $\beta_{y,z}$ can be reduced to a
one-dimensional integral, resulting in
\begin{eqnarray}
\beta_y =&   -{Z_P^2 \over \pi w v^2 b^2}
{\cal P} \int_{-\infty}^{\infty}d y {y^2 K_1^2(|y|) \over
                                        y - a}
                 \\
\beta_z =&  -{Z_P^2 \over \pi w v^2 b^2}
{\cal P} \int_{-\infty}^{\infty}d y {y^2 K_0^2(|y|) \over
                                        y - a}
\end{eqnarray}

It is noted  that,
 while $\alpha_{y,z}$ depends linearly on the projectile
charge $Z_P$,  $\beta_{y,z}$  does  quadratically.   By expanding
the
transition amplitude  in the  parameter  $Z_P$,
we actually get the
perturbative expansion Eq. 9.
Doing the same for a generic transition, we obtain
easily  the  matrices  ${\bf  V}^{(N-1)}$ and ${\bf  D}^{(N)}$\cite{note1}.
As the  procedure  is a long but  otherwise  straightforward  algebraic
task,  we  have  employed  a  commercial  code for algebraic  and  numerical
manipulation.

The development of
the  model  we  have  summarized  here  was motivated by two main  reasons.
First, it is exactly solvable so we can obtain
 both the exact  results and the
whole  perturbation  series  for  any  transition.    Second,  the  dipolar
approximation yields  a good representation  of atomic excitation in ion-atom
collisions for large values of the  projectile  charge  \cite{Rodriguez92}, the
better, the larger $Z_P$.  Thus,  departures of the model from a real atomic
collision system  arise  only from the target potential.

In  the  next  section, a comparison between the  exact  transition
amplitudes  and  the  different  perturbative  and  Schwinger  variational
results are presented.

\section{RESULTS AND DISCUSSION}

We are interested in representing the non perturbative region when $Z_P/v$
is greater than  one.    As  we will see, in this case the impact
parameter dependent
probabilities
exhibit  a maximum  for  increasing  impact parameters as $Z_P$ is
increased.   Hence, the dipole  approximation  is  consistent  with  this
feature, namely the importance of the  large  impact  parameters for strong
perturbations.  A similar behavior is actually  evidenced  in real atom
excitation by fast highly charged ion impact \cite{Rodriguez92}.
\begin{figure}
\includegraphics[width=8.13 cm, height=10.6 cm]{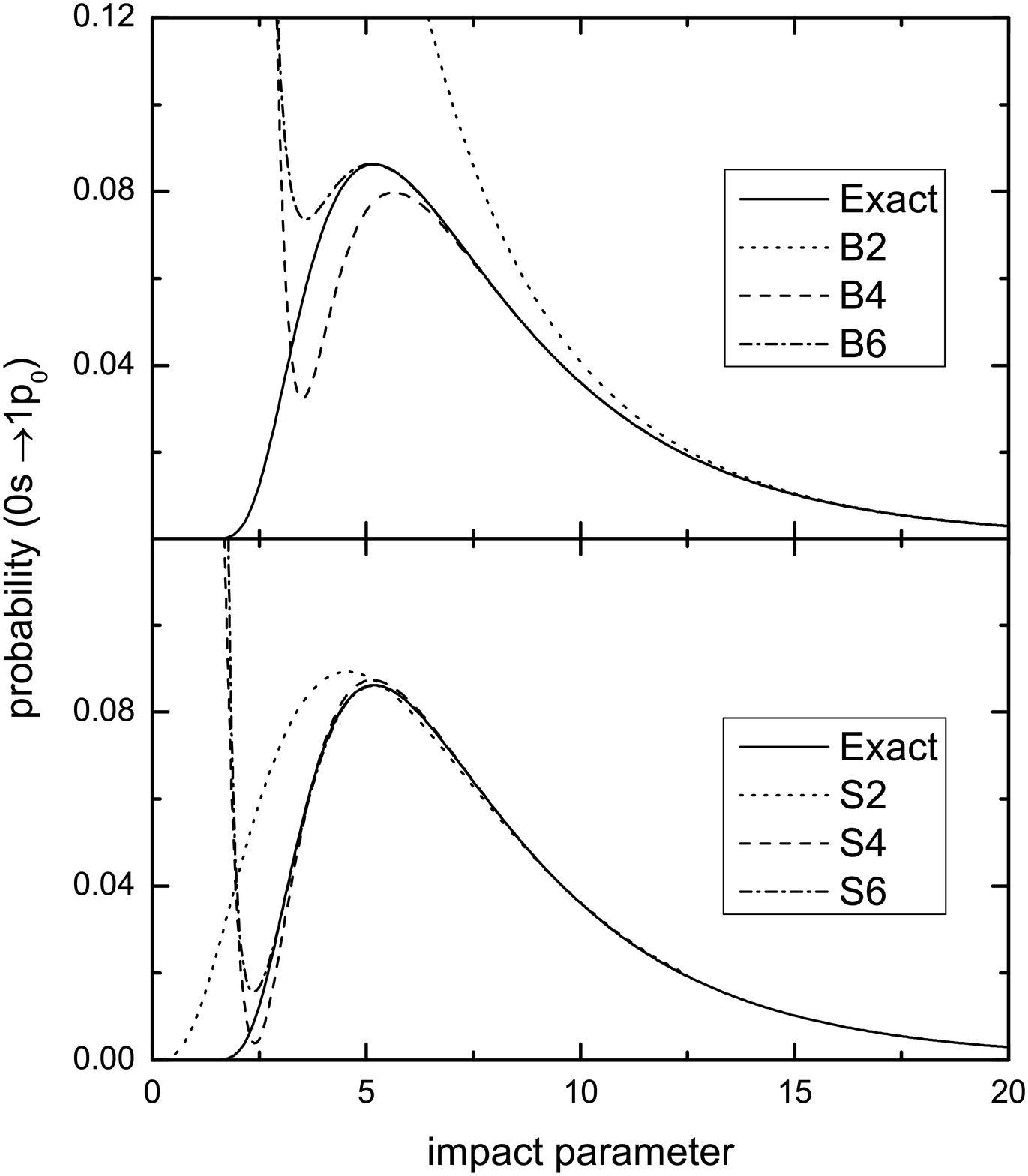}
\begin{caption}{Transition probabilities for the $\ket{0s}\to \ket{1p_0}$
transition as a function of the impact parameter for $Z_P=12$, $v=5$,
 and
 $w=0.5$. (a) Born approximation of order $n$, B$n$,
(b) Schwinger variational approximation of order $n$, S$n$ with 16 basis set expansion. Solid line curves correspond to exact model results.
}
\end{caption}
\label{Fig1}
\end{figure}

We intend to show the performance of the variational  principle by studying
two case studies
in detail.  The transition
probabilities as a function of the impact parameters
for a fixed $Z_P$, and
the  transition  probabilities  as  a function of the
impact parameters
for different
projectile charges $Z_P$, are presented.
 We fix the impact velocity in $v=5$,
and the oscillator  frequency  in  $w=0.5$.   It is an easy matter to show
that the present problem  possesses
  scaling properties relating the results for
different target frequencies $w$, and  also for different velocities.


   We  study  excitation  from the ground state $\ket{0s}$ to the
   first excited states $\ket{1p_0}$ and  $\ket{1p_1}$. We also discuss results for the elastic probabilities. The final states
   with well defined angular momentum can be
 obtained by using straight and well known transformation rules in terms of
the one-dimensional transition amplitudes
$a(\alpha,\beta,n,m)$.

\subsection{Fixed projectile charge as a function of the
impact parameter}
 Here, we analyze  the case where the  projectile  charge  is  fixed
 ($Z_P=12$), and the transition probabilities determined as a function  of
 the impact parameter.
 This is the normal procedure in order  to  obtain a
 total cross section.  The transition to a strong perturbation
 regime for a fixed $Z_P$ is  accounted for by  decreasing  the  impact  parameter.  We are
  interested in the  region  of impact parameter where the dipolar hypothesis
 remains valid, say $b>2$.   Indeed, as will be deduced below, this range is
 the one that mostly contributes  to  the  total  cross section as $Z_P$ becomes larger than $v$ in atomic units.
The calculations performed with the Born and with
the  Schwinger variational functional of order $N$ are
denoted by B$N$ and S$N$, respectively.
 For the trial initial and final wave functions
 we consider a basis set of the  16 lower energies states.

 First  transition probabilities for  $0s\to 1p_0$ (figure 1) and
  $0s\to 1p_1$ (figure 2) as a function of the impact parameter $b$ are analyzed.
In figures 1a and 2a
results  obtained with the Born series B2, B4 and B6 are displayed.
 It may be observed that the Born expansion remains
  close to the exact results
 for large   $b$.
 For the Born results of
  order 6, the results agree for $b>5$. For lower
  orders of the Born series, the exact
  results are only well represented for higher impact parameters.
  This behavior delimits  convergence properties
  of the Born series.

On the other hand, the corresponding results obtained
with the associated Schwinger functionals
are shown in figures 1b and 2b. The improvement over
the Born series is noteworthy. Even with the Schwinger functional
 of order
2 (S2),  the qualitative behavior is fairly improved.
 Notably,  the  S6 results are in good agreement with the exact ones
 for $b>2.5$. The Schwinger functionals are convergent as the
 order is increased, although for small impact parameters
 the probabilities are largely overestimated by the higher orders. This is a well known
 problem with variationally obtained magnitudes when the trial wave functions are not  reasonably close to the exact ones.

 Briefly, as the order of both the Schwinger and the Born amplitudes increases the results approaches the exact ones. However, as expected, Schwinger variational calculation
 are a good improve over Born ones.
  \begin{figure}
\includegraphics[width=8.13 cm, height=10.6 cm]{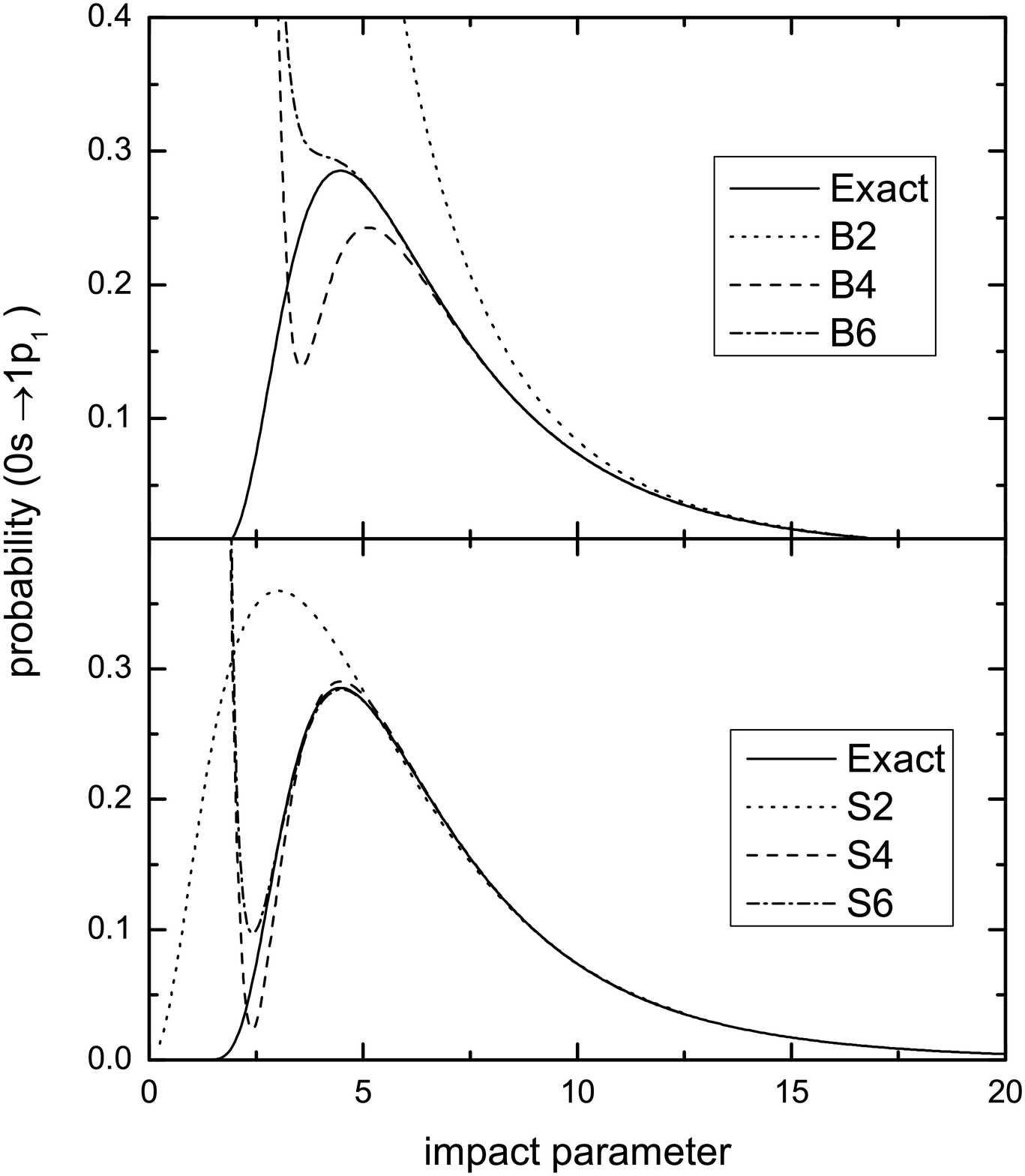}
\begin{caption}{Transition probabilities for the $\ket{0s}\to \ket{1p_1}$
transition as a function of the impact parameter for $Z_P=12$, $v=5$,
 and
 $w=0.5$. (a) Born approximation of order $n$, B$n$,
(b) Schwinger variational approximation of order $n$, S$n$ with 16 basis set expansion.  Solid line curves correspond to exact model results.
}
\end{caption}
\label{Fig1}
\end{figure}


 \subsection{Probabilities for different projectile charges}
In the results obtained for
 different projectile charges as a function of the
  impact parameter,
 the transit to a strong  perturbation  regime is taken into account
by increasing  the  projectile charge.
  This analysis  is motivated  by  previous
 ideas \cite{Rodriguez92}  concerning  the  behavior  of  the transition amplitudes for
 large values of the projectile charges.
  First Born approximations give a
 simple $Z_P^2$ scaling  for  the  transition  probabilities,  {\it  i.e.},
 probabilities do increase in  a  monotonous way with the projectile charge.

 \begin{figure}
\includegraphics[width=8.13 cm, height=10.6 cm]{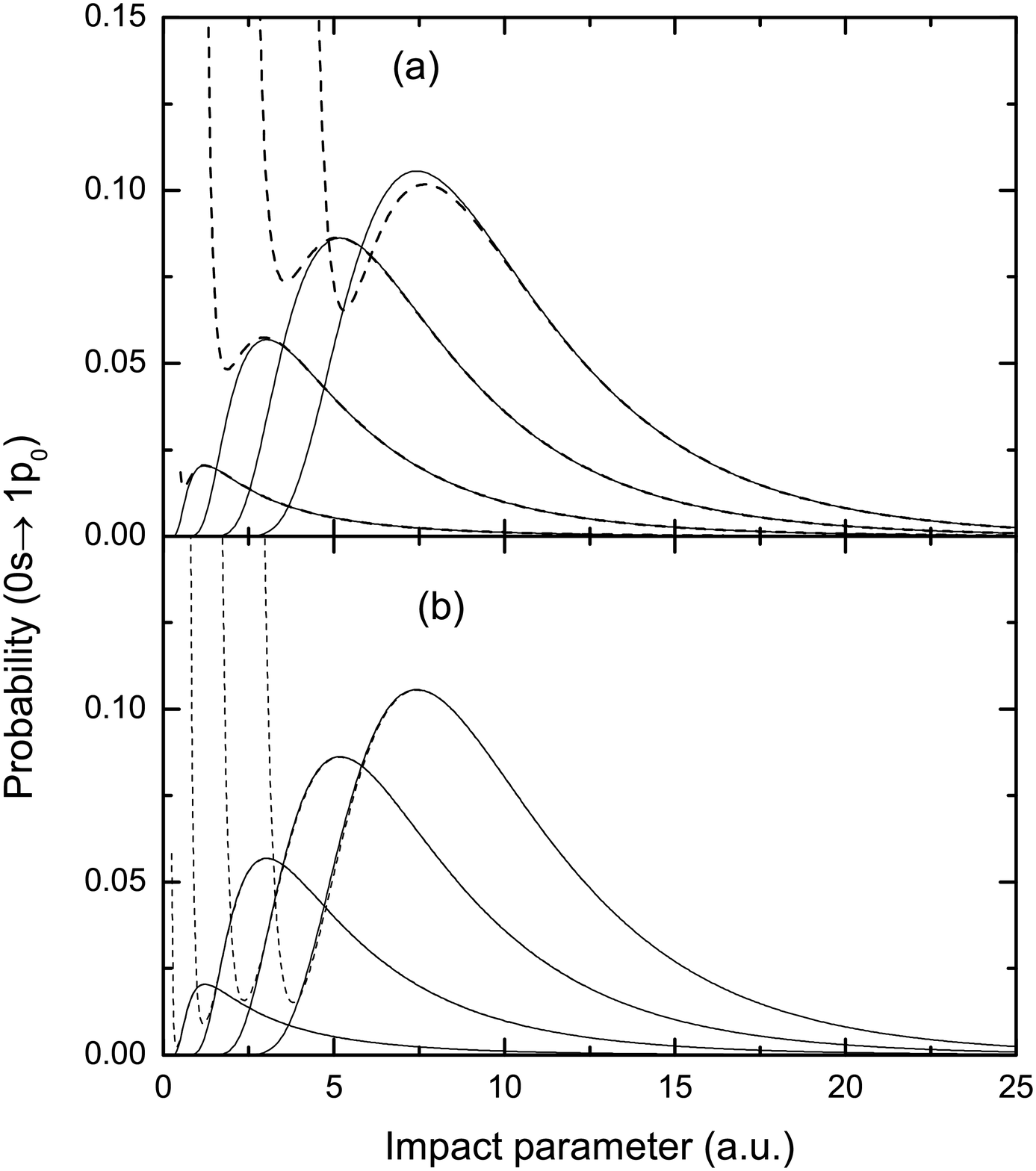}
\begin{caption}{Transition probabilities for the $\ket{0s}\to \ket{1p_0}$
transition as a function of the impact parameter for different projectile charges $Z_P=$ 2, 6, 12, and 20
($w=0.5, ~ v=5$). (a) Born approximation of order $6$, B$6$,
(b) Schwinger variational approximation of order $6$, S$6$ with 16 basis set expansion.
 Solid line curves correspond to exact model results. For increasing value of $Z_P$
 the  maxima in the curves  are shifted towards larger impact parameter.
}
\end{caption}
\label{Fig3}
\end{figure}
\begin{figure}
\includegraphics[width=8.13 cm, height=10.6 cm]{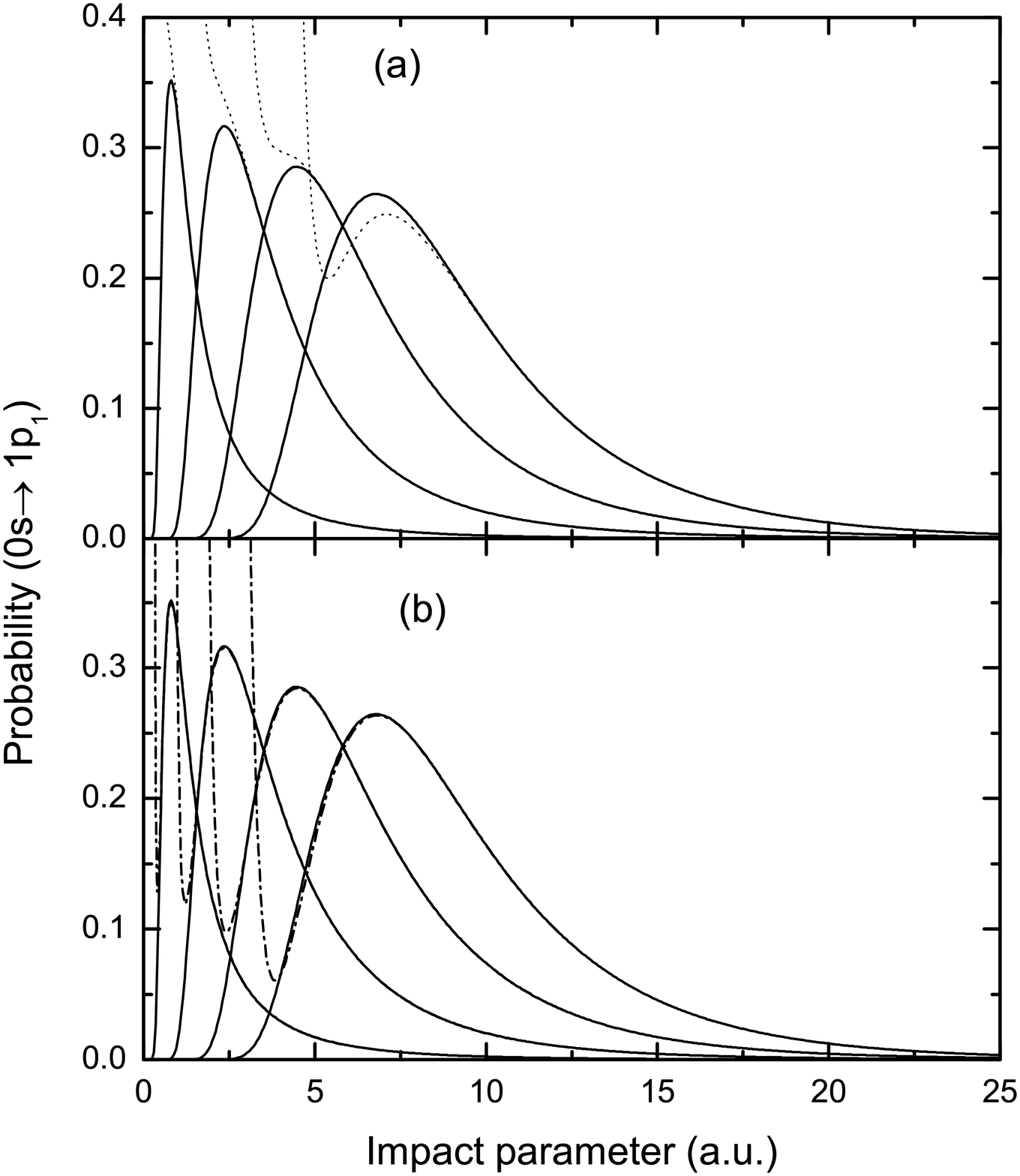}
\begin{caption}{As in figure 3 but for the $\ket{0s}\to \ket{1p_1}$ transition probabilities.
}
\end{caption}
\label{Fig4}
\end{figure}

 Transition probabilities for  $0s\to 1p_0$ (figure 3) and
  $0s\to 1p_1$ (figure 4) transitions as a function of the impact parameter $b$ are show again. Four values of the projectile charge: $2,6,12$ and $20$ are considered.
 Now, we focus on a comparison between the results Born results B6
(figure 3a an 4a ) with the Schwinger S6 ones (figure 3b and 4b).
 Probabilities show
  maxima at increasing  impact parameter as the projectile charge is
  increased, following roughly a $\sqrt{Z_P}$ law. Thus, for higher charges  the distant collisions yield the
  larger transition probabilities. Therefore distant collisions contributes the most to  total cross sections.
    Probabilities  obtained with the simplest  B6 agree with the exact only beyond the maxima curves. For  $0s\to 1p_1$ transition not even the maxima are displayed by this high order Born approximation calculations.
 In the case of   the S6 results, the agreement with the exact
 curves is both qualitative and  quantitative for
 $b>4$ in the case of $Z_P=20$, and for $b>1$ in the case of $Z_P=2$.
 For lower values of $b$, the Schwinger results break down, and
 probabilities are again largely overestimated. However as mentioned before  these impact parameters range are not relevant 
 total cross sections.

  \subsection{Convergence with the basis expansion set}
  \begin{figure}
\includegraphics[width=8.94  cm, height=6.1 cm]{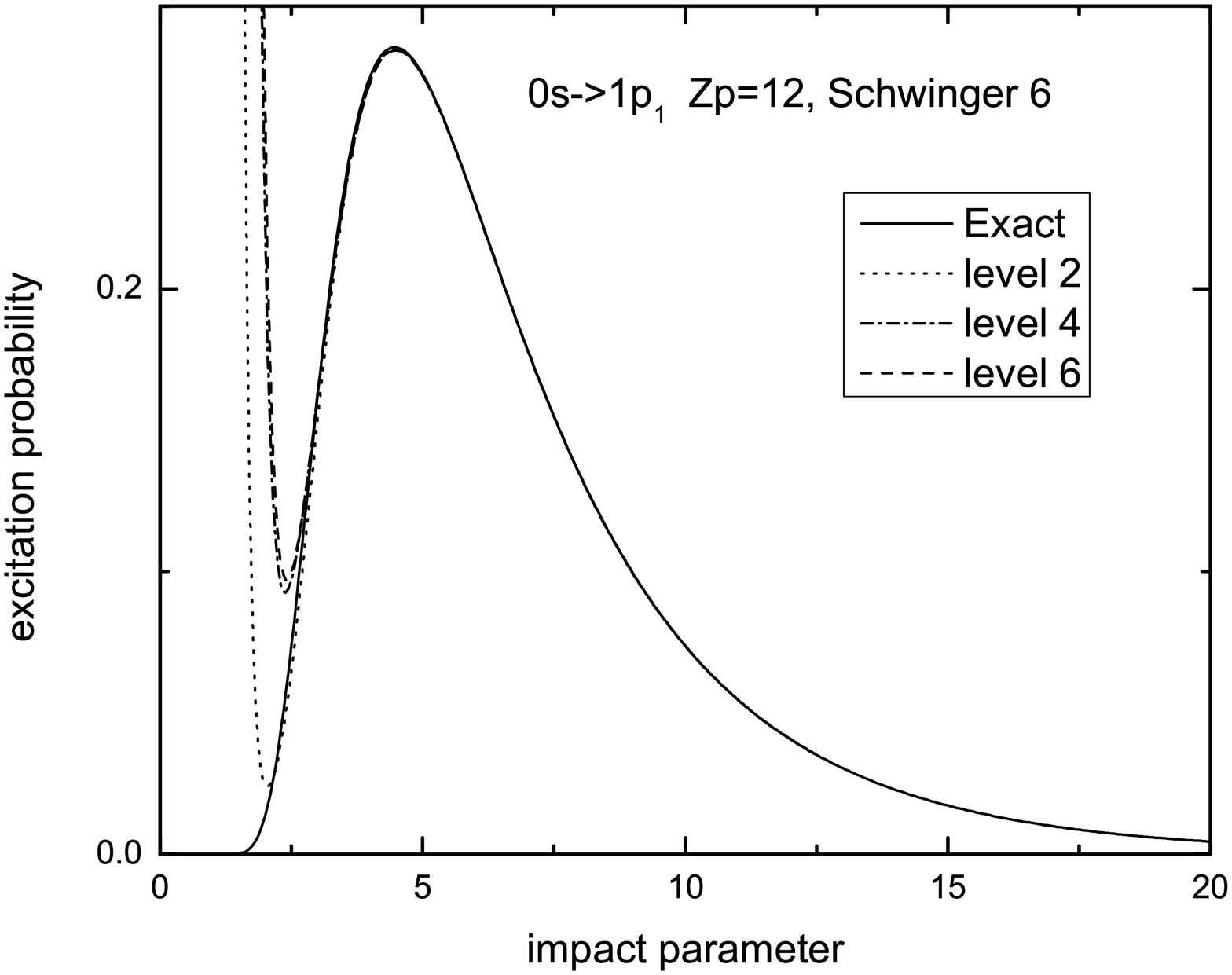}
\begin{caption}{ Probabilities for the $\ket{0s}\to \ket{0s}$
elastic process as a function of the impact parameter for $Z_P=12$, $v=5$,  and
$w=0.5$.
 Schwinger variational approximation of order $6$ (S$6$)
 for different number of states in the basis set.
 Solid curves are exact results.
}
\end{caption}
\label{Fig5}
\end{figure}
 
Finally,
results obtained with the Schwinger
functional of order 6, but for different
sizes of basis sets,
are displayed in figure 5. The elastic probabilities are shown. The  S6 fails when
only  the
initial and  final  states are included in the basis set.
 In this case, the same results
as the  Born approximation of order 2 are obtained. For elastic probabilities and accounting for 
 selection rules, the results with a single state basis set gives unity for any impact parameter.
   Moderate to large basis states are in good agreement with the exact results.
 The limit case of a very large basis set is well represented by
 using 16 states.  No further improvement is obtained by increasing
 the basis set over this size.

   The ultimate reason for this kind flat behavior is that the coefficients in the trial wave functions are 
   required to be independent of time. If the coefficients would be allowed to have a time dependence 
   the description would become accurate for any impact parameter as the basis set increases.

\section{conclusions}

Martin and Salin \cite{Martin95}
 have reported
 some doubts about the convergence properties of the
 Born series
 in ion-atom excitation problems
 based on close-coupling calculations.
 From the present work, we
conclude that  convergence  of  the  Born series is assured for large
 impact parameters.  As the Born order is increased, convergence
for decreasing impact parameters is obtained. It
looks like that for  large
enough   Born series order 
  the minimum impact parameter that is well described may be pushed down
even more.

 The improvement obtained by using
 the Schwinger  functionals  arises from increasing both, the order and the
 basis set size.
    In  this  work,  we  have  shown  the superiority of the
 variational calculations as compared with Born results of the same order.
 We remark that, once the Schwinger convergence fails,
 the probabilities are even worse than those obtained
 with the Born series.  The
performance of the Schwinger functional depends critically on the accuracy
of the Born terms. This is particularly true for large
projectile charges because of the multiplication of the
$N$-order term by $Z_P^N$.
 Wether this level of convergence of the Schwinger
variational amplitudes may be  obtained  in  the  real  ion-atom  collision
problem is an open question.    However,  this  work
may  encourage further research to address this line.


\begin{references}
\bibitem{note1} Note that the  phases $\beta_{y,z}$ should be properly taken
into account in the expansion.
\end{references}
\end{document}